\documentclass[cits,preprint]{PoS}

\usepackage{url}
\usepackage{amsmath}
\usepackage{mathbbol}

\newcommand{\bra}[1]{\langle{#1}|}
\newcommand{\ket}[1]{|{#1}\rangle}

\title{Meson and baryon masses with low mode averaging}

\ShortTitle{Meson and baryon masses with low mode averaging}

\author{Gunnar~Bali, \speaker{Luca~Castagnini}, Sara~Collins\\
         Institut f\"ur Theoretische Physik, Universit\"at Regensburg, \\
93040 Regensburg, Germany\\
        E-mail: \email{gunnar.bali@physik.uni-regensburg.de}, \email{luca.castagnini@physik.uni-regensburg.de}, \email{sara.collins@physik.uni-regensburg.de} }

\abstract{We describe and test a method known in the literature as low mode averaging
to improve Euclidean two-point functions in lattice QCD using the
low-lying eigenmodes of the Wilson-Dirac operator $D$. The contribution from the
low modes is averaged over all positions of the quark sources while the contribution
from high modes is calculated in the  traditional way using one source point per lattice.
We apply this method to different baryon and meson two-point functions and we compare the 
improvements using the eigenmodes of the non-hermitian operator $D$ and the eigenmodes
of the hermitian operator $Q=\gamma_5 D$. The convergence strongly depends on the parity
of the states.
}

\FullConference{The XXVIII International Symposium on Lattice Field Theory, Lattice2010\\
		June 14-19, 2010\\
		Villasimius, Italy}

\begin{document}

\section{Introduction}
Solving the Dirac equation $D \psi = \eta$ may be a very time consuming
task at small quark masses as the condition number of $D$ is proportional
to the inverse of the quark mass. Some approaches
involve the computation
of the low-lying eigenvectors of $\gamma_5D$ and use these to deflate the Dirac
operator,  see e.g.\ ref.~\cite{Morgan:2004,Stathopoulos:2007zi}. These eigenvectors
can also be used to reduce the noise of the signal with a technique
known in the literature \cite{DeGrand:2004qw,Giusti:2004yp} as low mode
averaging (LMA). This consists of decomposing the quark propagator
into a sum of high and low modes $\psi=\psi_{\mathrm{high}}+\psi_{\mathrm{low}}$
and in averaging
the low mode contribution over all lattice points. We define the
LMA two-point function as,
\begin{equation}
C_{\mathrm{LMA}} (t) = C_{\mathrm{low}}(t) + C^{\mathrm{pa}}(t) - C^{\mathrm{pa}}_{\mathrm{low}}(t)\,,
\end{equation} 
where $C^{\mathrm{pa}}$ is the exact point-to-all correlation function, calculated
for a single source point. This definition satisfies
$\langle C_{\mathrm{LMA}} (t) \rangle = \langle C^{\mathrm{pa}} (t) \rangle$, however,
the errorbars are reduced, due to more sampling per lattice.
In the present work we apply the LMA technique to meson and baryon
 two-point functions with degenerate quark masses. We study the
improvement as a test for future work on hadron spectroscopy
and three-point functions.

\section{Mesons}
The present study is based on 100 configurations with
lattice volume $V=16^3\times32$ generated with the quenched
Wilson action at $\beta=6.0175$ using
Chroma \cite{Edwards:2004sx}. This
corresponds to a lattice spacing of $a\approx0.2093\, \sigma^{-1/2} \approx
0.093\,\mathrm{fm}$. For each configuration we computed the lowest $30$
eigenvectors of the massive hermitian Dirac operator
$\gamma_5 D |u_i\rangle = \lambda_i | u_i\rangle $ using the algorithm
by Kalkreuter and Simma \cite{Kalkreuter:1995mm} at  $\kappa=0.1557$
corresponding to $m_\pi \approx 425$~MeV and $m_{\pi}L\approx 3.2$.
We can reconstruct the contribution of the low modes to the quark propagator
as~\cite{Neff:2001zr}, $D^{-1}_{\mathrm{low}}  = \sum_i \frac{1}{\lambda_i}   \ket{u_i}\bra{u_i} \gamma_5$ while $C_{\mathrm{low}}(t)$ is given by,
\begin{equation}
\label{eq:gammastruc}
C_{\mathrm{low}} (t) = \sum_{i,j,x, \mathbf{y}}\frac{1}{\lambda_i \, \lambda_j}  \bra{ u_j({x})}  \gamma_5 \Gamma \, \ket{u_i(x)} \bra{u_i({y})} \gamma_5 \Gamma \, \ket{u_j ({y})}\,, 
\end{equation} 
where $y_4=x_4+t$.

Clearly the LMA technique works best when the low modes are dominant.
This happens for the $\pi$ as can be seen
from fig.~\ref{fig:a0Corr} (left) where we compare the smeared-smeared
$\pi$ point-to-all correlation function with the low mode
contribution. However, for positive parity mesons 
(see fig.~\ref{fig:a0Corr} (right) for the $a_0$),
the low modes saturate the two-point function very slowly.
This can also be seen by comparing the LMA and point-to-all
smeared-smeared effective masses 
of fig.~\ref{fig:pieff} for negative parity mesons with the
positive parity ones of fig.~\ref{fig:a1eff}.
In the case of the $a_0$ the low mode contribution even has a wrong curvature
in the central region, see fig.\ \ref{fig:a0Corr} (right).

The meson correlators differ in the $\Gamma$ structure of the interpolating fields:
for the $\pi$ (the best case) we have a total $\Gamma$ product
$\gamma_5 \Gamma = \mathbb{1}$ in eq.~(\ref{eq:gammastruc}) while for the
$a_0$ this reads  $\gamma_5 \Gamma = \gamma_5$. One way to get rid of a $\gamma_5$ factor
is to calculate the right and left eigenvectors of the non-hermitian Dirac operator,
$ D |r_i\rangle = \lambda_i | r_i\rangle$, $\bra{\ell_i}D = \lambda_i \bra{\ell_i} $
and to construct the quark propagator as \cite{Bali:2009hu,Guerrero:2010bf},
$D^{-1}_{\mathrm{low}} =  \sum_i \frac{1}{\lambda_i}   \ket{r_i}\bra{\ell_i}$. In this way the
meson two-point function becomes,
\FIGURE[h]{\vspace{-8pt}
\includegraphics[width=0.48\textwidth,clip]{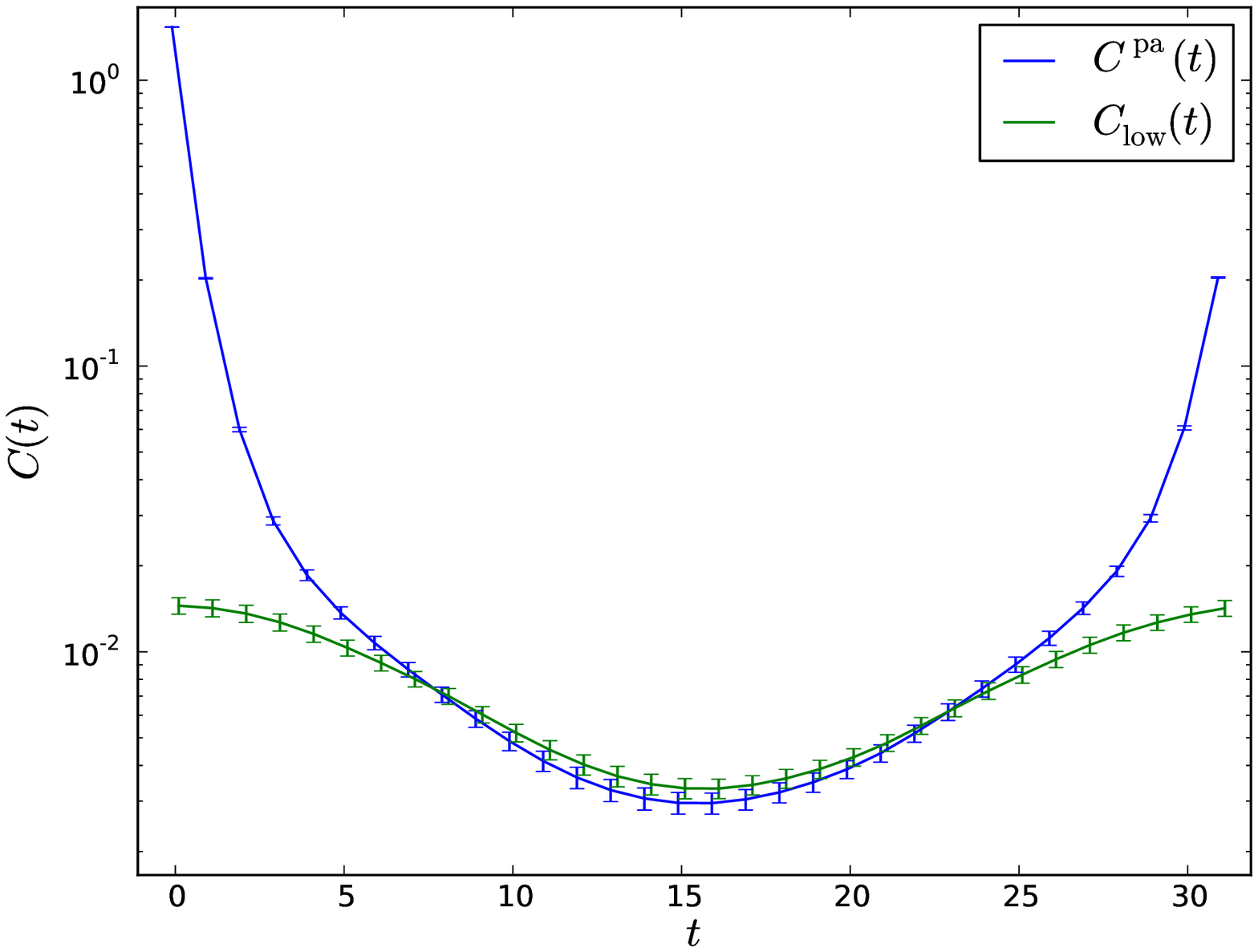}\hspace*{.03\textwidth}
\includegraphics[width=0.48\textwidth,clip]{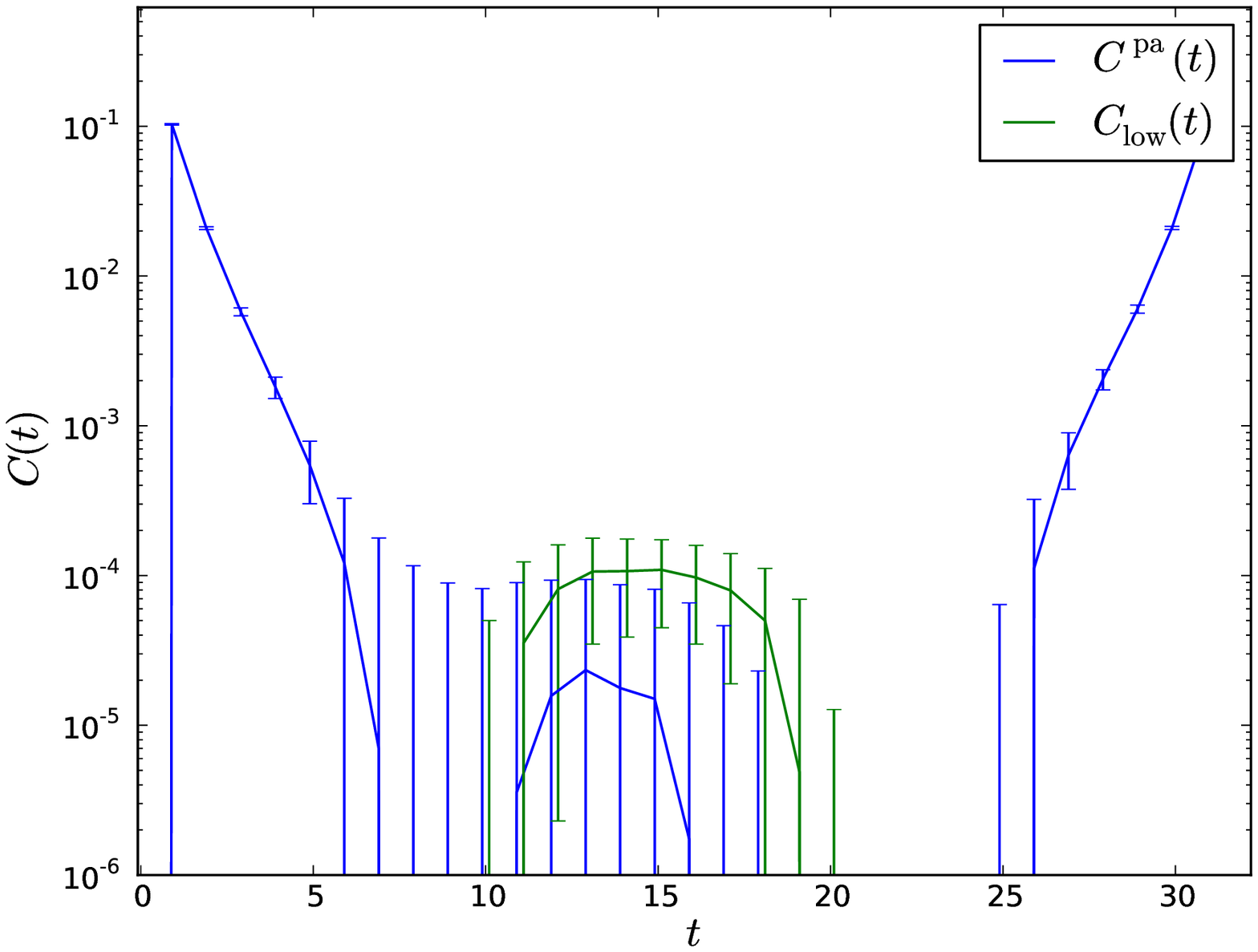} 
 \vspace{-20pt}
\caption{Low mode saturation (30 modes): $\pi$ ($J^{PC}=0^{-+}$, left)
and $a_0$ ($J^{PC}=0^{++}$, right)
two-point functions.\label{fig:a0Corr}}}
\FIGURE[h]{
\includegraphics[width=0.48\textwidth,clip]{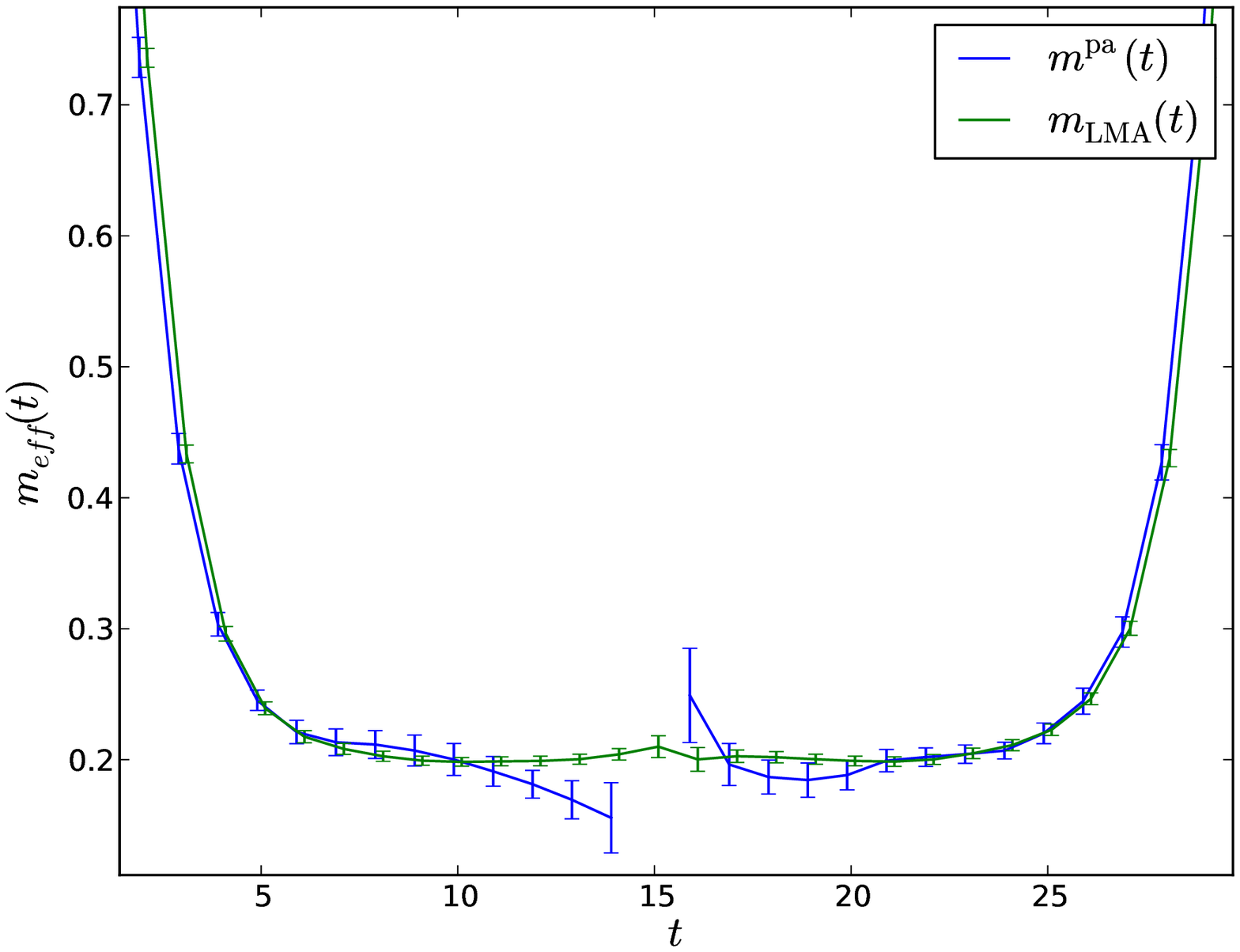}\hspace*{.03\textwidth}
\includegraphics[width=0.48\textwidth,clip]{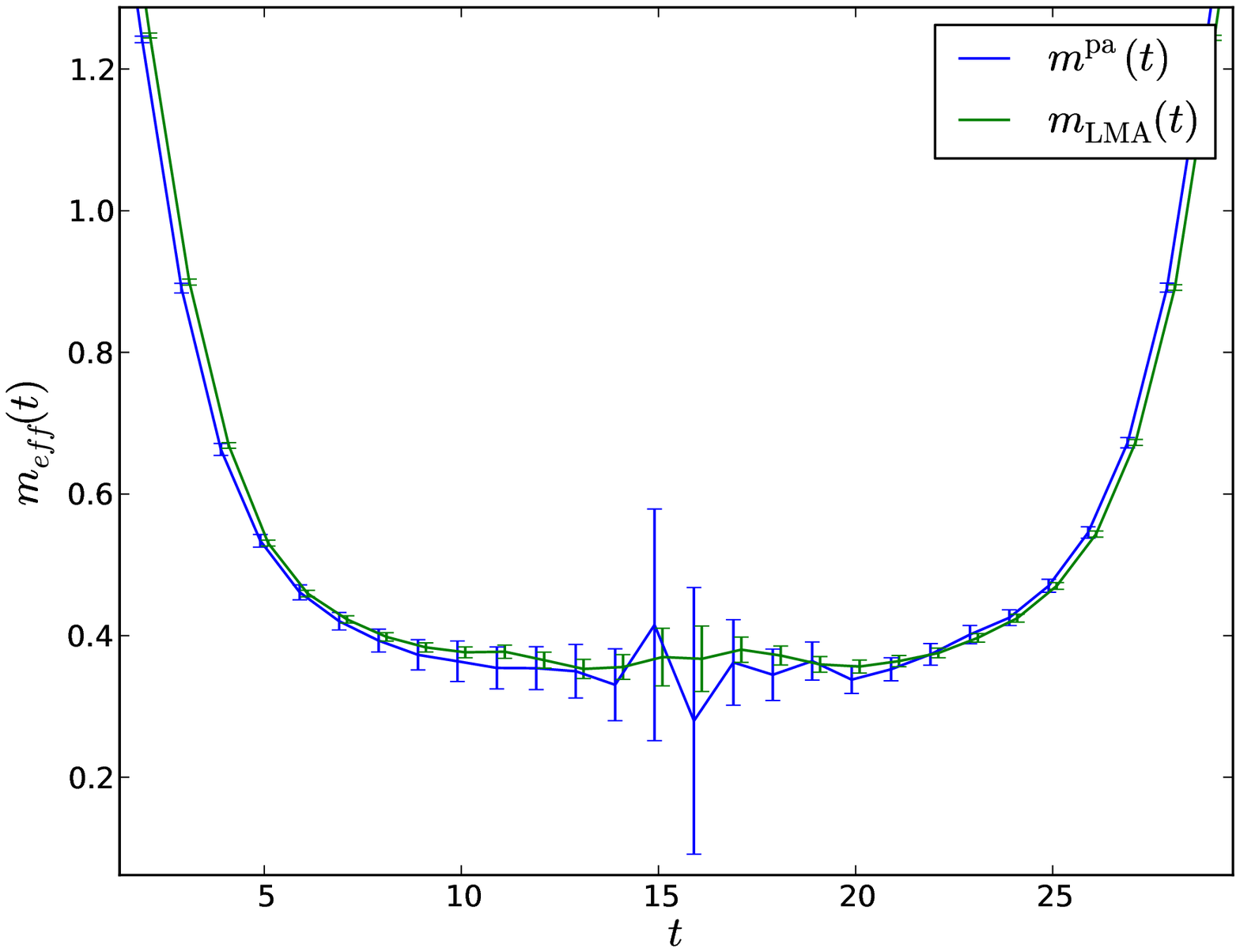}
 \vspace{-20pt}
\caption{Effective masses: $\pi$ ($J^{PC}=0^{-+}$, left) and
$\rho$ ($J^{PC}=1^{--}$, right)\label{fig:pieff}}}
\FIGURE[h!]{
\includegraphics[width=0.48\textwidth,clip]{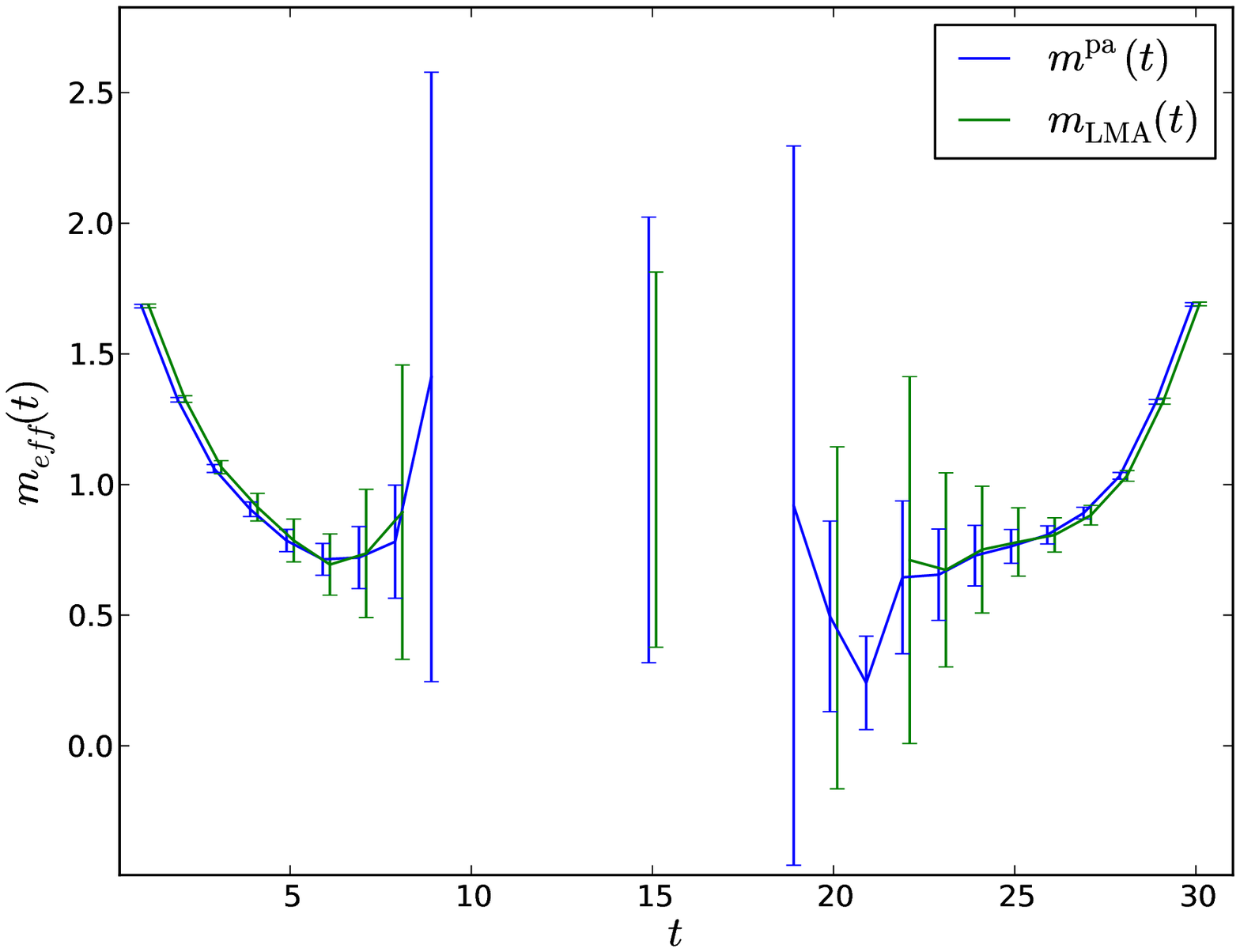}\hspace*{.03\textwidth}
\includegraphics[width=0.48\textwidth,clip]{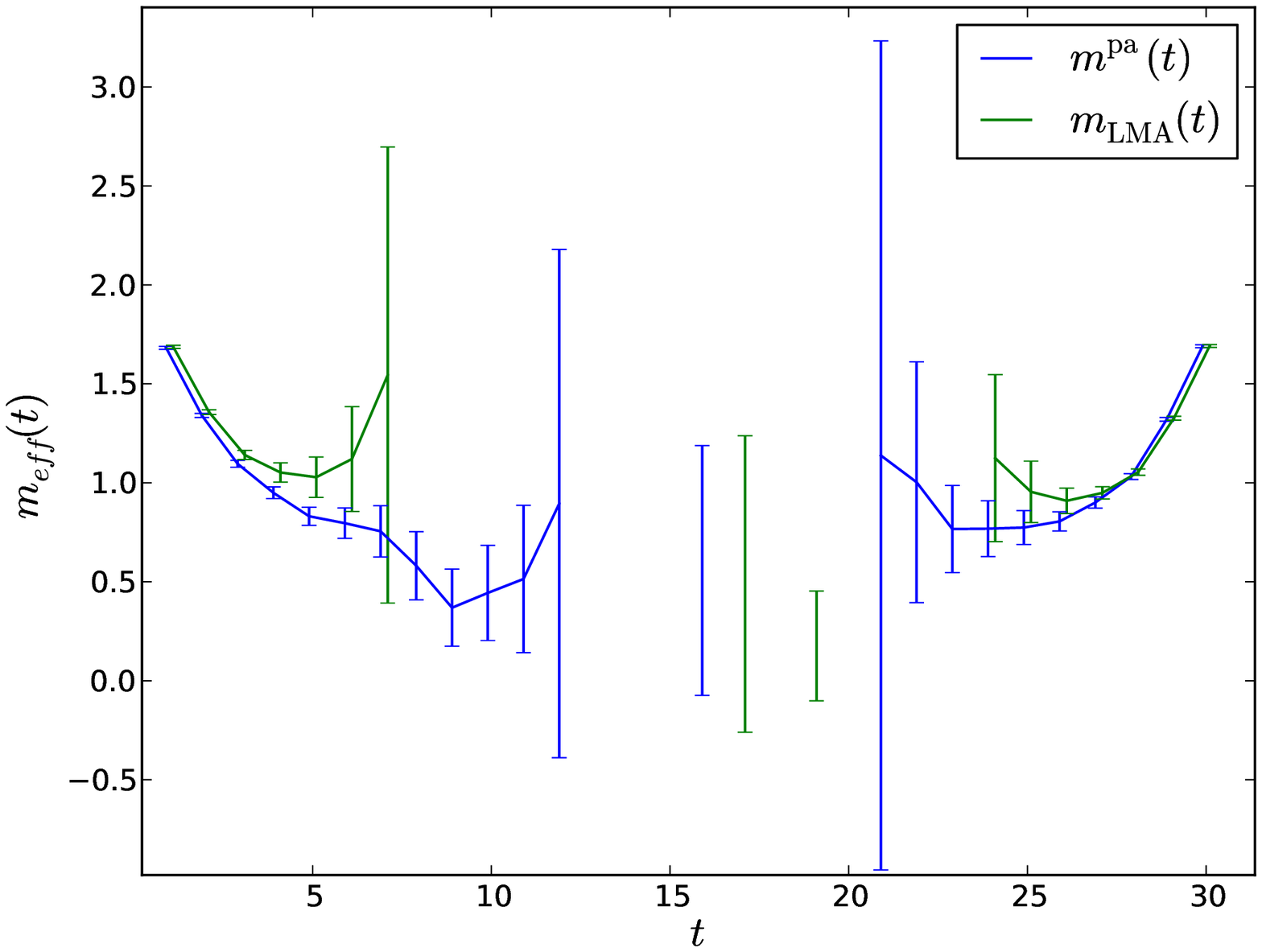}
\vspace{-20pt}
\caption{Effective masses: $a_1$ ($J^{PC}=1^{++}$, left) and
$b_1$ ($J^{PC}=1^{+-}$, right).\label{fig:a1eff}}}
\begin{equation}
C_{\mathrm{low}}(t) =  \sum_{i,j,x, \mathbf{y}}\frac{1}{\lambda_i \, \lambda_j}  \bra{ \ell_j({x})}  \Gamma \, \ket{r_i(x)} \bra{ \ell_i({y})} \Gamma \, \ket{r_j ({y})}\,.
\end{equation} 
Note that $\langle\ell_i|r_j\rangle= \delta_{ij}$ and
it can easily be seen that due to the property
$D^{\dagger}=\gamma_5D\gamma_5$,
$\lambda_i^*$ is an eigenvalue whenever
$\lambda_i$ is an eigenvalue,
with a left eigenvector $\langle \tilde{\ell}_i|=\langle r_i|\gamma_5$ and
a right eigenvector $|\tilde{r}_i\rangle=\gamma_5|\ell_i\rangle$.
\TABULAR[h!]{|c|c|c|}{
\hline &  $m $& $m_{\mathrm{LMA}}$ \\ \hline
 $\pi$ & $0.2003(49)$ & $0.2019(36)$ \\ \hline
 $\rho$ & $0.3965(66)$ & $0.3952(41)$ \\ \hline}{
Fitted meson masses in lattice units.\label{tab:meson} }

We used the Arnoldi method implemented in the ARPACK library to compute the
non-hermitian eigenvectors. This method is much slower than that of Kalkreuter and Simma.
However, the eigenvectors are independent of the hopping parameter $\kappa$ and
when changing the quark mass we only need to rescale the eigenvalues $\lambda$.
Unfortunately, we could not find any significant improvement using 15 pairs of
non-hermitian low modes over the standard point-to-all method.
This may be related to
the fact that the ratios $|\lambda_{30}|/|\lambda_{1}|$ were found to be
by factors of approximately five smaller in the non-hermitian case than for the hermitian case.

From the fit to the correlators we found a 30\% improvement
on the extrapolated $\pi$ and $\rho$ meson masses for the hermitian LMA,
as displayed in table~\ref{tab:meson}.
\FIGURE{
\includegraphics[width=0.48\textwidth,clip]{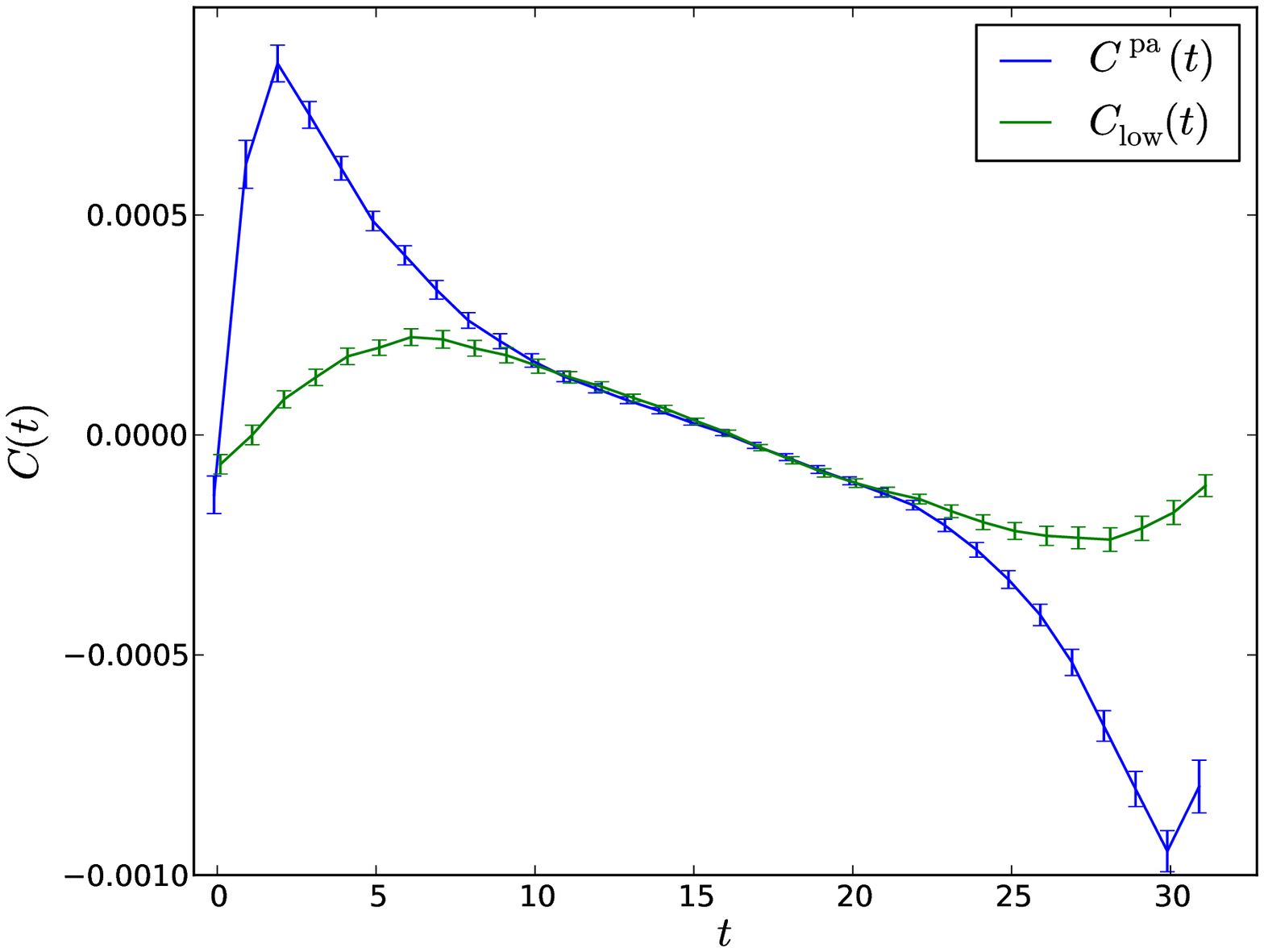} \vspace{-10pt}
\caption{Local axial-vector current.\label{fig:axial}}}
\noindent We also applied the method to the local axial-vector current
(fig.~\ref{fig:axial}), to test the improvement on the pion decay constant
$f_\pi$. Following the conventions of ref.~\cite{Gockeler:1997fn} we set,
\begin{equation}
 m_\pi \, f_\pi =\langle 0 | \gamma_4\gamma_5 |\pi \rangle =- 2 \kappa \frac{\sqrt{2 m_\pi} A_{A P}}{\sqrt{A_{P P}}}\,.
\end{equation}

We denote the amplitudes of correlation functions with
a smeared pseudoscalar sink and a local axial-vector source
as $A_{A P}$ and with a local pseudoscalar source as $A_{P P}$.
The improvement of LMA is very striking in this case,
\begin{equation}
   f_\pi = 0.079(27)\,,\quad f_{\pi, \mathrm{LMA}} = 0.082(12)
\,.
 \end{equation}

\section{Baryons}
The low mode averaging technique for baryons is more challenging due to the way
the interpolators are contracted. 
For the nucleon $N$, the $\Delta^+$ and the $\Lambda$ we use the following interpolators,
\begin{align}
N(x) &=  \varepsilon_{abc} u(x)_a \left(u(x)_b^T C \gamma_5 d(x)_c \right)\,,  \\
\Delta^+(x) &=\varepsilon_{abc}\left [  2 \left(u(x)_a^T C \gamma_\mu d(x)_b \right) u(x)_c +  \left(u(x)_a^T C \gamma_\mu u(x)_b \right) d(x)_c   \right ]\,,\\
\Lambda(x) &= \varepsilon_{abc}\left [  2 \left(u(x)_a^T C \gamma_5 d(x)_b \right) s(x)_c +  \left(u(x)_a^T C \gamma_5 s(x)_b \right) d(x)_c -  \left(d(x)_a^T C \gamma_5 s(x)_b \right) u(x)_c      \right ] \,,
\end{align}
where $C$ is the charge conjugation operator.
Note that we only study the mass-degenerate case,
$m_u=m_d=m_s$. In the following, we consider
the contractions for the example of the
nucleon two-point function,
\begin{equation}
  \langle N (y)\overline{N}(x)   \rangle =  \langle \varepsilon_{abc}  \varepsilon_{a'b'c'} (C \gamma_5)_{\alpha \beta} (C \gamma_5)_{\alpha'\beta'} (P_\pm )_{\gamma \gamma'} d(y)_{b' \beta'} \overline {d}(x)_{b \beta}   u(y)_{a' \alpha'} \overline {u}(x)_{a \alpha} u(y)_{c' \gamma'} \overline {u}(x)_{c \gamma}  \rangle\,,\label{eq:contra}
\end{equation}
where $P_\pm =\frac{1}{2} (\mathbb{1}\pm \gamma_4)$ is the parity projector.

The expression eq.~(\ref{eq:contra}) can be written
in terms of the left and right eigenvector components
in a way that the source term is decoupled from the sink term:
\begin{equation}
 \begin{split}   \langle N(y) \overline{N}(x)   \rangle = & \sum_{i,j,k} \frac{1}{\lambda_i\,\lambda_j \,\lambda_k}   \left\{ \varepsilon_{a'b'c'} (C \gamma_5)_{\alpha' \beta'} \left[ r_{i,\, b',\, \beta'}(y)  r_{j,\, a',\, \alpha'}(y)  r_{k,\, c',\, \gamma'}(y) \right] \right\}  (P_\pm )_{\gamma \gamma'}\\
& \times  \left\{  \varepsilon_{abc}  (C \gamma_5)_{\alpha \beta} \ell^*_{i,\, b,\, \beta} (x)\left[ \ell^*_{j,\, a,\, \alpha} (x) \ell^*_{k,\, c,\, \gamma} (x) - \ell^*_{j,\, c,\, \gamma} (x)\ell^*_{k,\, a,\, \alpha} (x)\right] \right\} \,.
 \end {split}
\end{equation}
For the hermitian case, $\gamma_5D|u_i\rangle=\lambda_i|u_i\rangle$, $|r_i\rangle=|u_i\rangle$
and $\langle\ell_i|=\langle u_i|\gamma_5$.
The cost of computing the above expression increases with the third power of the number of
eigenmodes used.
To optimize this computation, we split up each eigenvector $r$ into twelve $r_{a,\alpha}$ spin-colour
components and store them as lattice objects. Multiplications
like $r_{i,\, a,\, \alpha}  r_{j,\, b,\, \beta}$ are optimized in QDP++.
With these and other optimizations the cost of the
contractions for 30 eigenmodes becomes
negligible relative to the cost of solving for the propagator.

\FIGURE{\vspace{-8pt}
\includegraphics[width=0.48\textwidth,clip]{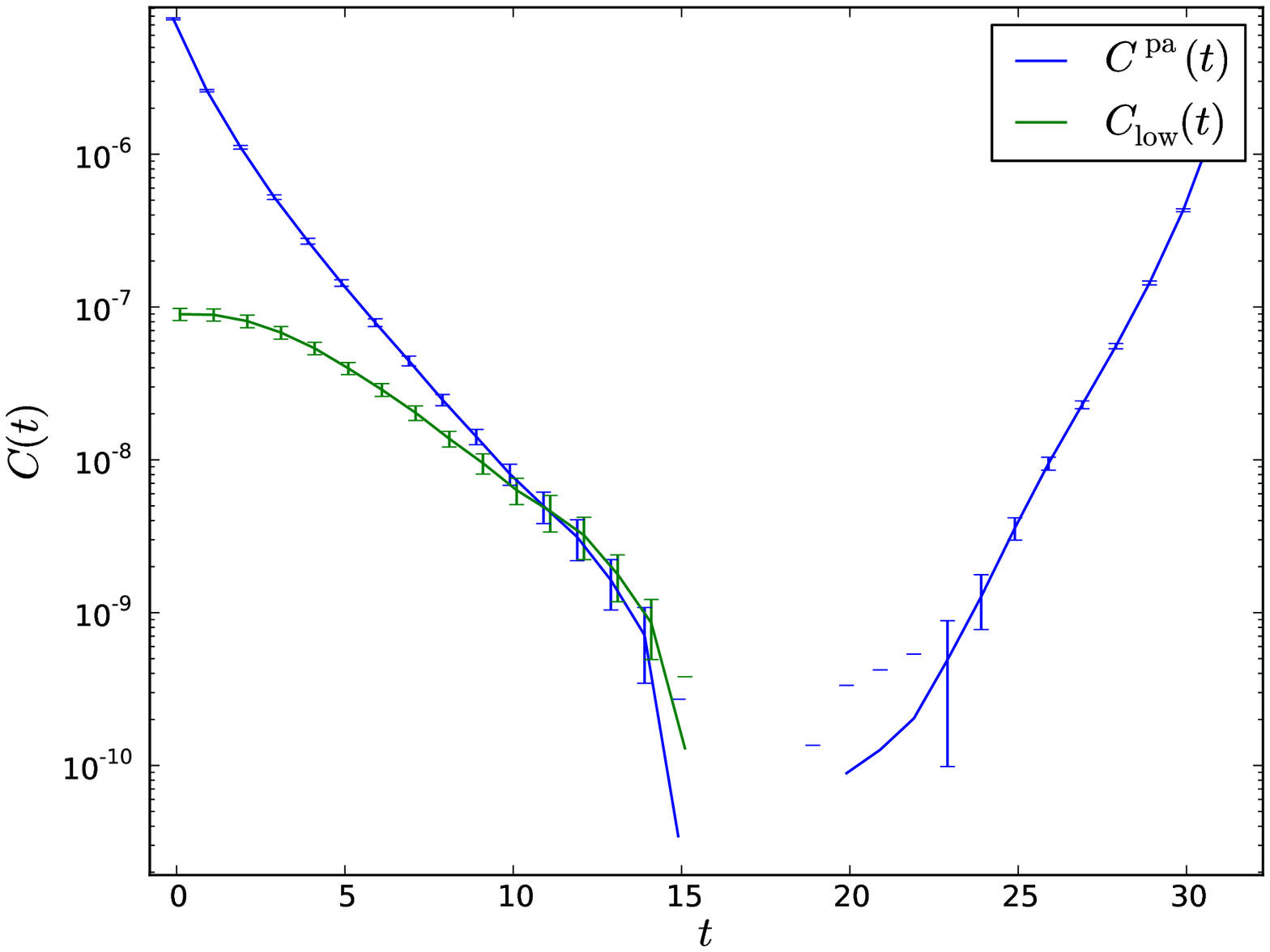}\hspace*{.03\textwidth}
\includegraphics[width=0.48\textwidth,clip]{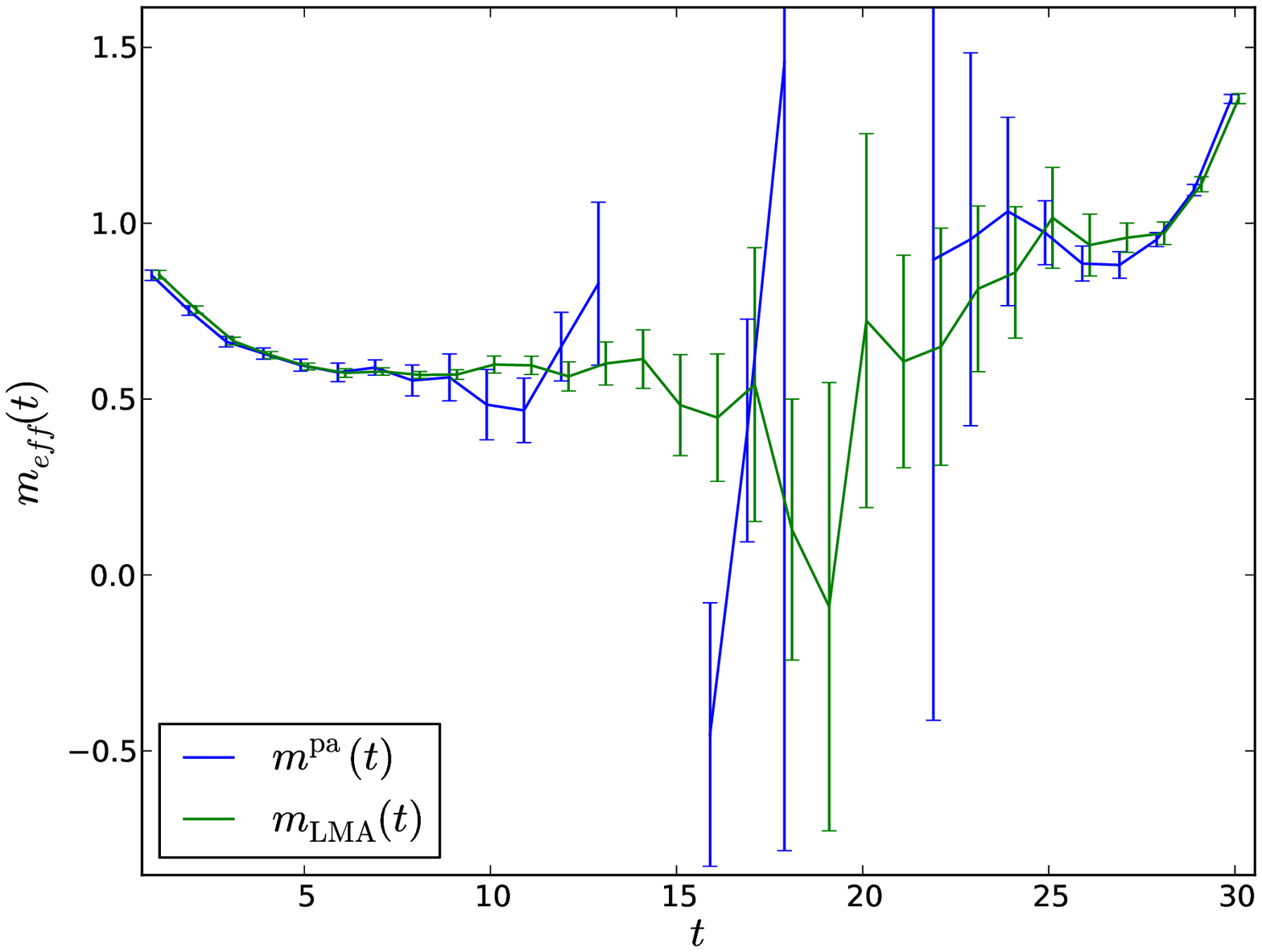}\vspace{-20pt}
\caption{Nucleon ($J^P=\frac{1}{2}^+$) two-point function and effective masses.
The backwards propagating state is the $N^*$ ($J^P=\frac{1}{2}^-$).
\label{fig:proton}}}

\FIGURE{\vspace{-15pt}
\includegraphics[width=0.48\textwidth,clip]{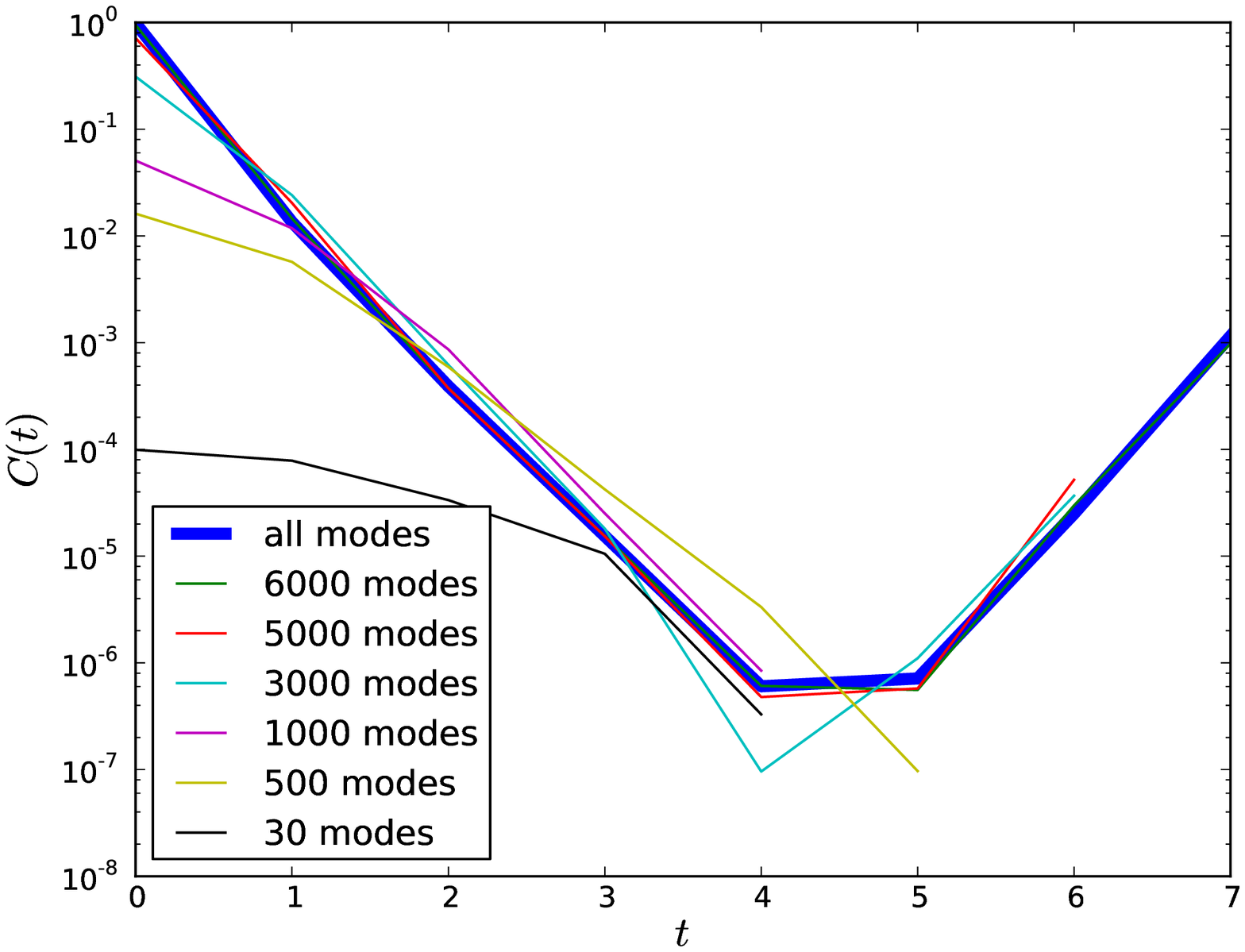}\hspace*{.03\textwidth}
\includegraphics[width=0.48\textwidth,clip]{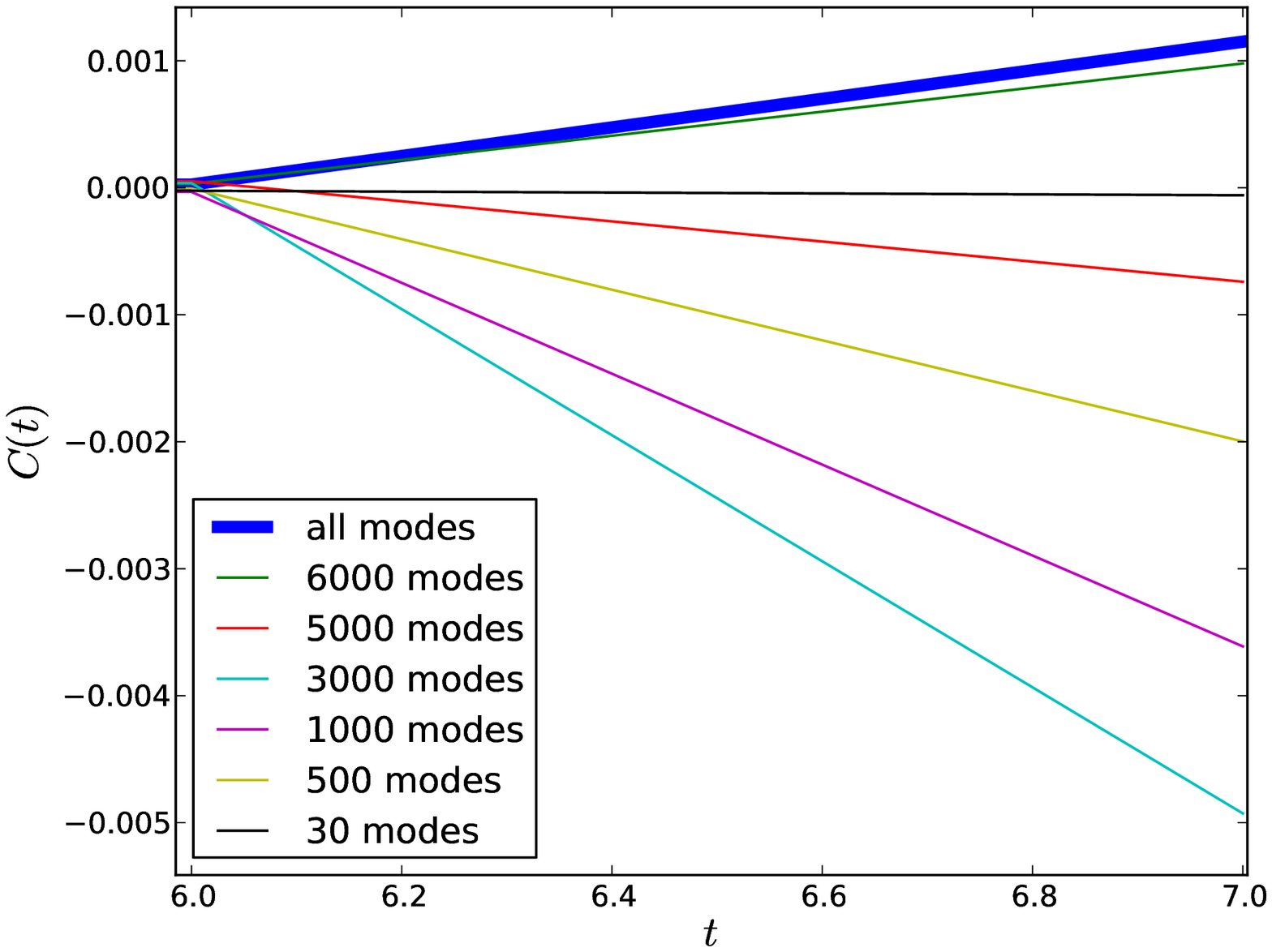}\vspace{-20pt}
\caption{Low mode contribution to the nucleon two-point function on
a $4^3\times 8$ lattice. On the right hand side the $t=6,7$ data
are shown without the logarithmic scale.
\label{fig:feweigs}}}

In fig.~\ref{fig:proton} we
show the low mode saturation of the nucleon two-point function
and the LMA improvement of the corresponding effective mass.
The low mode contribution for the $N^*$ turns out to be negative
and is not visible on the logarithmic scale.
To gain a better understanding of this behaviour we investigated a small 
$4^3\times 8$ lattice
volume where we calculated all the $6144$ eigenvectors of the Wilson operator
using the LAPACK library. 
In fig.~\ref{fig:feweigs} we display the nucleon two-point function
$C_{\mathrm{low}}(t)$ on such a single
configuration for different numbers of eigenmodes. For the positive parity state 
that propagates from $t=0$ into the forward direction the low modes are dominant and
quickly saturate the correct correlation function. Conversely, for the backward
propagating states even the sign is wrong until over 85~\% of the modes are summed up.
Eventually, after all modes are included, the correct result is obtained.
We also tested non-hermitian LMA for baryons. This did not solve the sign problem
for negative parity states and, as in the mesonic case, no improvement over the
conventional point-to-all method was found.

\FIGURE{
\includegraphics[width=0.48\textwidth,clip]{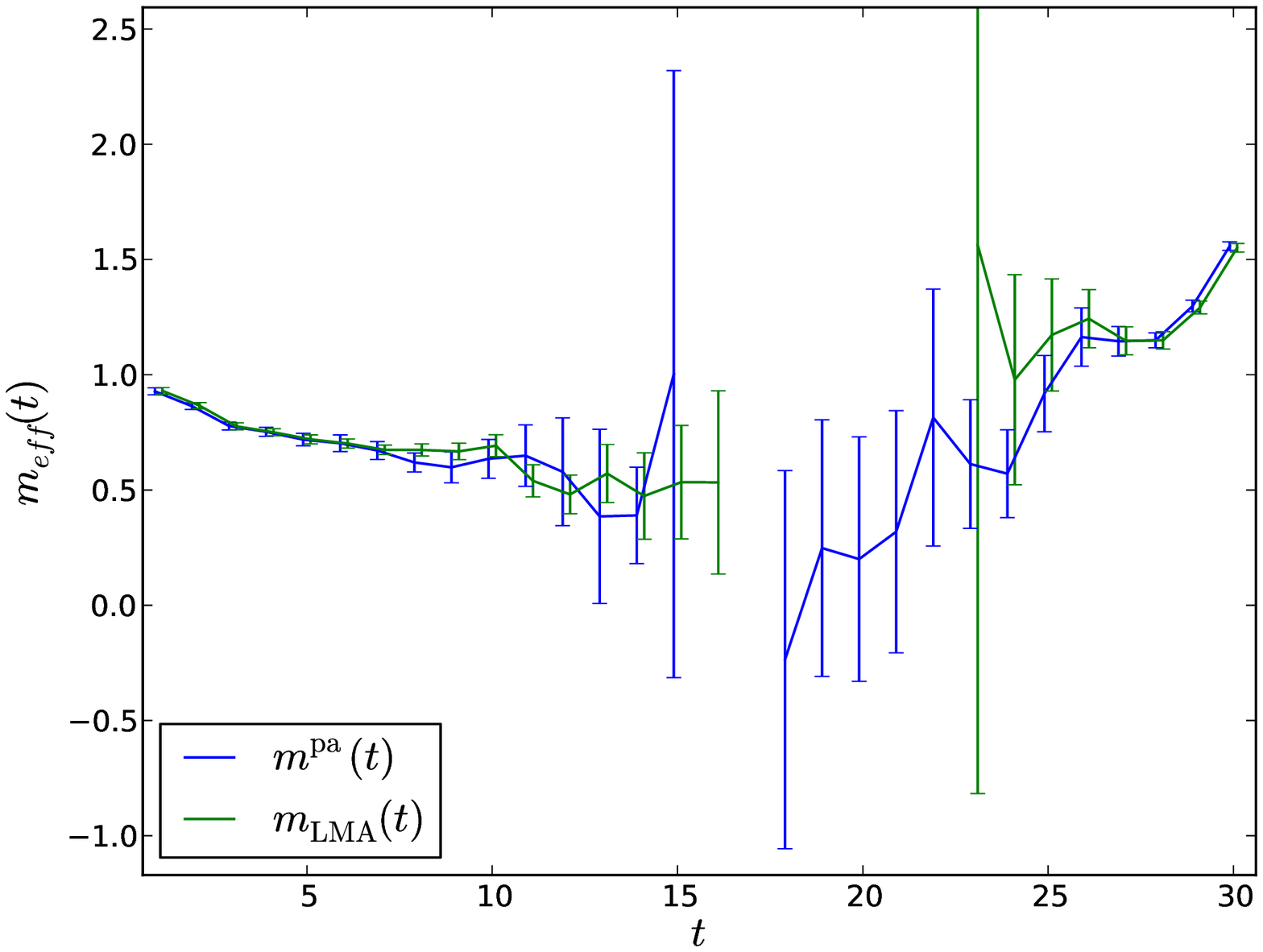}\hspace*{.03\textwidth}
\includegraphics[width=0.48\textwidth,clip]{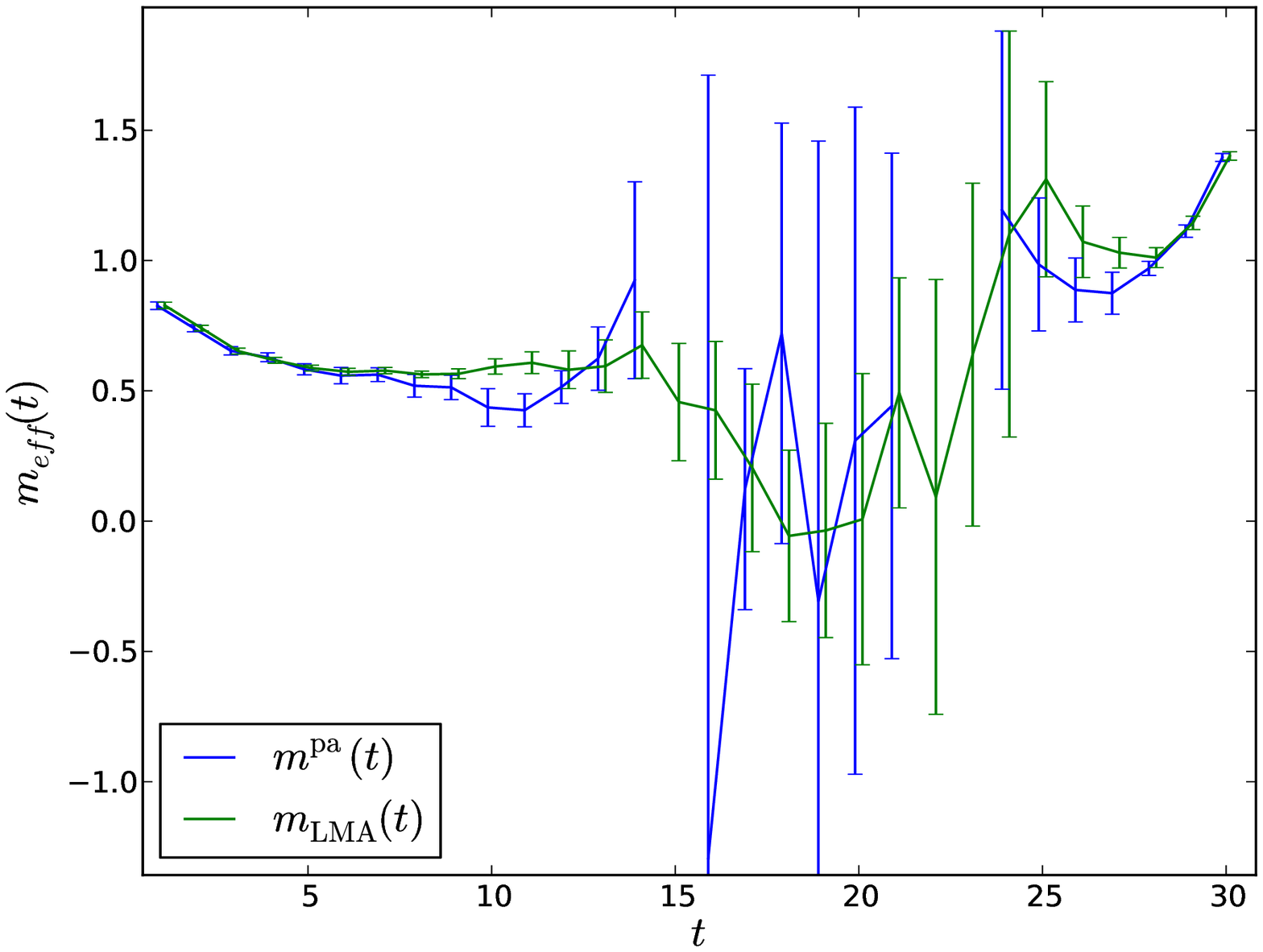}\vspace{-15pt}
\caption{Effective masses for the $\Delta^+$ ($J^P=\frac{3}{2}^+$, left)
and $\Lambda$ ($J^P=\frac{1}{2}^-$, right).
\label{fig:deltalambda}}}

\TABULAR{|c|c|c|}{
 \hline &  $m $& $m_{\mathrm{LMA}}$ \\ \hline
 $p$ & $0.604(16)$ & $0.595 (7)$\\ \hline
 $\Delta^+$ & $0.717(20)$ & $0.713(12)$ \\ \hline
 $\Lambda$ &$0.586(17)$ & $0.592(9)$\\ \hline}{
Fitted baryon masses in lattice units
for $m_u=m_d=m_s$.\label{tab:baryon}}

For the $\Delta^+$ and the $\Lambda$ a behaviour similar to that of the nucleon was found.
In fig.~\ref{fig:deltalambda} we display the effective mass plots.
The negative parity partners are not well approximated by the lowest modes that exhibit
the wrong sign.
However, LMA reduces the statistical errors for the forward propagating states.
The errors on the fitted positive parity baryon masses are
reduced by factors of roughly two, using the hermitian LMA method,
see table~\ref{tab:baryon}.

\section{Conclusions}
We confirm that the hermitian low mode averaging technique is an efficient method to reduce
the noise of two-point functions and of the fitted hadron masses. It works well
for negative parity mesons and very well for positive parity baryons but it fails completely
for the opposite parity cases where low mode saturation does not set in at all. Non-hermitian
low mode averaging was inefficient for all particles studied, due to its slower
saturation behaviour. 

It should be pointed out that computing the low lying eigenmodes
is computationally demanding.
Preliminary tests on a $24^3\times 48$ volume with $m_{\pi}\approx 420$~MeV
($m_{\pi} L\approx 4.8$) show that it is more efficient 
to increase the number of point sources
than to compute the eigenvectors, to reduce the errors on meson masses.
For baryons the LMA is cost-efficient. If one further
decreases the quark mass then on one hand the costs of computing
propagators will drastically
increase while on the other hand low mode saturation may set in faster,
making this method even
more efficient.
Moreover, eigenvectors may be recycled for deflating the
Dirac operator, in particular if multiple source points per
configuration are used,
see e.g.\ refs.~\cite{Morgan:2004,Stathopoulos:2007zi}. In this
case the overhead of LMA is negligible.

\acknowledgments
We thank Jacques Bloch for discussions. The simulations were run on
Regensburg's Athene HPC cluster. Sara Collins was supported by the
Claussen-Simon-Foundation (Stifterverband f\"ur die deutsche Wissenschaft).
We acknowledge support from Deutsche Forschungsgemeinschaft (Sonderforschungsbereich/Transregio 55) and the European Union grant 238353 (ITN STRONGnet).

\end{document}